\documentclass[aps,pre,twocolumn,superscriptaddress,showpacs]{revtex4-1}
\usepackage{amssymb}
\usepackage{natbib}
\usepackage[english]{babel}
\usepackage[dvips]{graphics}
\usepackage{graphicx,epsfig}
\usepackage{amsmath}
\usepackage{amsfonts}
\usepackage{color}
\usepackage{multirow}
\usepackage[normalem]{ulem}
\usepackage{booktabs}
\usepackage{soul}
\usepackage{lineno}

\newcommand{\limfunc}[1]{\textnormal{#1}} 
\newcommand{\func}[1]{\textnormal{#1}}      
\begin{document}


\title{Towards AC-induced optimum control of dynamical localization}
\author{F. Revuelta}
\affiliation{Grupo de Sistemas Complejos,
  Escuela T\'ecnica Superior de Ingenieros Agr\'onomos,
  Universidad Polit\'ecnica de Madrid,
  Avda.\ Complutense s/n 28040 Madrid, Spain.} 
\affiliation{Instituto de Ciencias Matem\'aticas (ICMAT), 
  Cantoblanco, 28049  Madrid, Spain.}

\author{R. Chac\'{o}n}
\affiliation{Departamento de F\'{\i}sica Aplicada, E.\ I.\ I., 
  Universidad de Extremadura, Apartado Postal 382, 06006 Badajoz, Spain}
\affiliation{Instituto de Computaci\'on Cient\'ifica Avanzada,
  Universidad de Extremadura, Apartado Postal 382, 06006 Badajoz, Spain}

\author{F. Borondo}
\affiliation{Instituto de Ciencias Matem\'aticas (ICMAT), 
  Cantoblanco, 28049  Madrid, Spain.} 
\affiliation{Departamento de Qu\'imica, 
  Universidad Aut\'onoma de Madrid, Cantoblanco, 28049 Madrid, Spain.}

\date{\today}

\begin{abstract}
It is shown that optimum control of dynamical localization 
(quantum suppression of classical diffusion) 
in the context of ultracold atoms in periodically shaken optical 
lattices subjected to time-periodic forces having equidistant zeros 
depends on the \textit{impulse} transmitted by the external force 
over half-period rather than on the force amplitude. 
This result provides a useful principle for optimally controlling dynamical
localization in general periodic systems, which is capable of experimental
realization.
\end{abstract}

\maketitle

%
%

\section{Introduction}

\label{sec.intro}

Quantum effects in transport phenomena in classical systems represent an
interesting fundamental issue in quantum theory that started from a remark
of Einstein in his celebrated paper on torus quantization \cite{1}. 
One of such effects, widely studied in the context of time-periodic systems, 
is the quantum suppression of classical chaotic diffusion \cite{2,3,4,5,6,7}, 
or dynamical localization (DL) for short. 
Remarkably, this effect is a quantum manifestation of the fact that a 
time-periodic force can stabilize a system, and it is thus expected 
to play a key role in our understanding of the problem of quantum-classical
correspondence in classically chaotic systems \cite{8,9}.
 While it is natural to think that, with the period fixed, this effect must 
depend on the temporal rate at which energy is transferred from the driving 
mechanism to the system, i.e., on the force waveform, the main target of 
study up to now has only been their dependence on the force amplitude 
because of the traditional use of sinusoidal forces. 
Recent work has provided strong evidence for a different dependence 
of DL on sinusoidal and square-wave forces \cite{10}. 
Since there are infinitely many different waveforms, 
a natural question arises: How can the influence of the shape of a periodic 
force on DL be quantitatively characterized. 
In this work, we demonstrate that for space-periodic systems subjected 
to a generic AC time-periodic force with equidistant zeros such 
characterization is well provided by a single quantity: 
the \textit{impulse} transmitted by the force over half-period 
---hereafter called \textit{force impulse}. 
This impulse is a quantity that accounts simultaneously
for the force's amplitude, period and waveform.


The organization of the paper is as follows. 
In Sec.~\ref{sec.theory}, we describe our model system of ultracold atoms 
in a periodically shaken optical lattice as well as our theoretical 
approach to  characterize the effect of the force impulse on DL. 
Section~\ref{sec.results} provides numerical confirmation of the 
effectiveness of the force impulse according to the theoretical 
predictions of Sec.~\ref{sec.theory} as well as discussing its influence
on the quantum dynamics and underlying classical structures.
Finally, we conclude by presenting in Sec.~\ref{sec.conclu} 
the main conclusions of our work along with some final remarks.

\section{Theoretical approach}

\label{sec.theory}

\begin{figure}[tbp]
\includegraphics[width=0.6\columnwidth,angle=270]{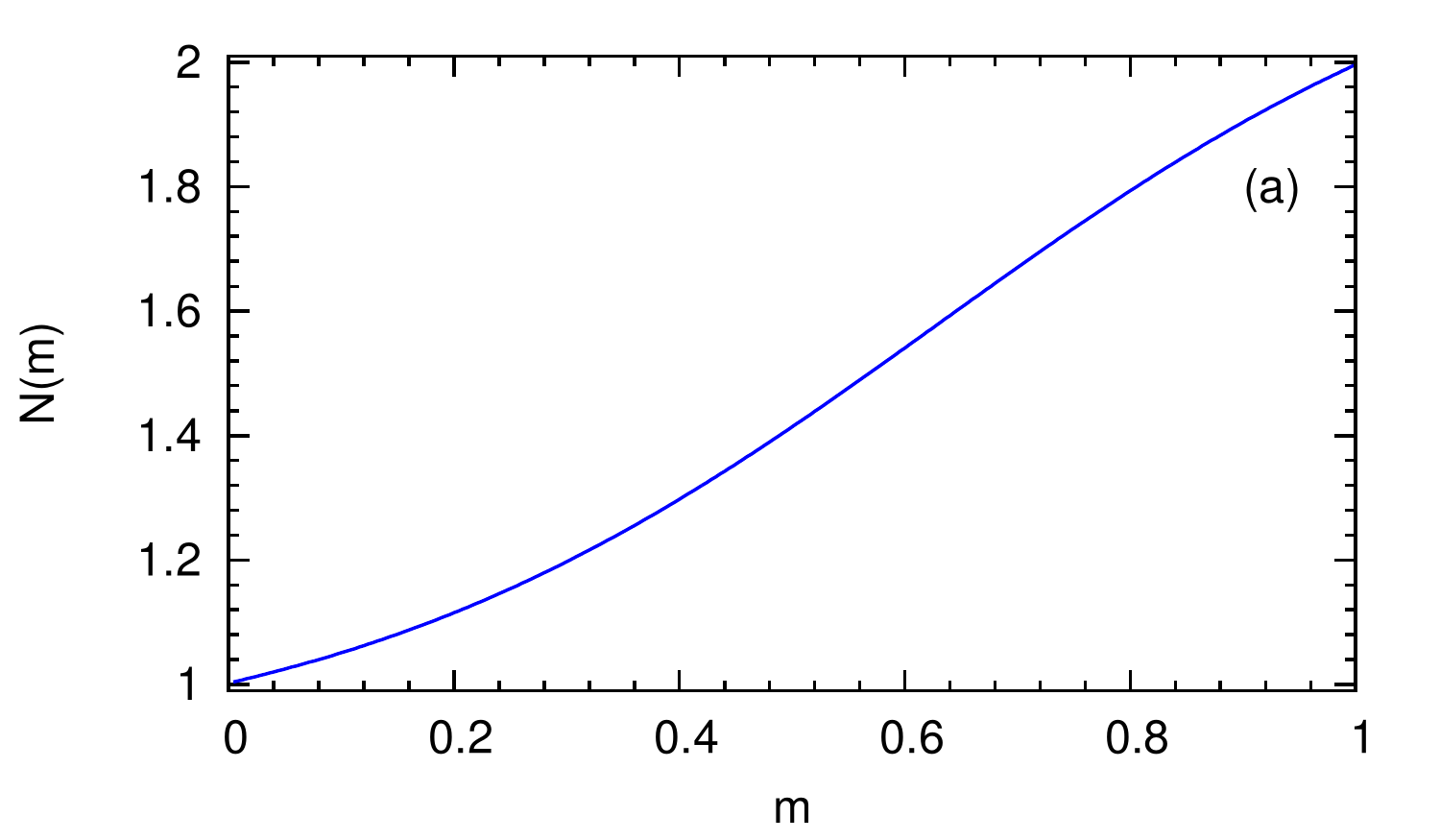} %
\includegraphics[width=0.6\columnwidth,angle=270]{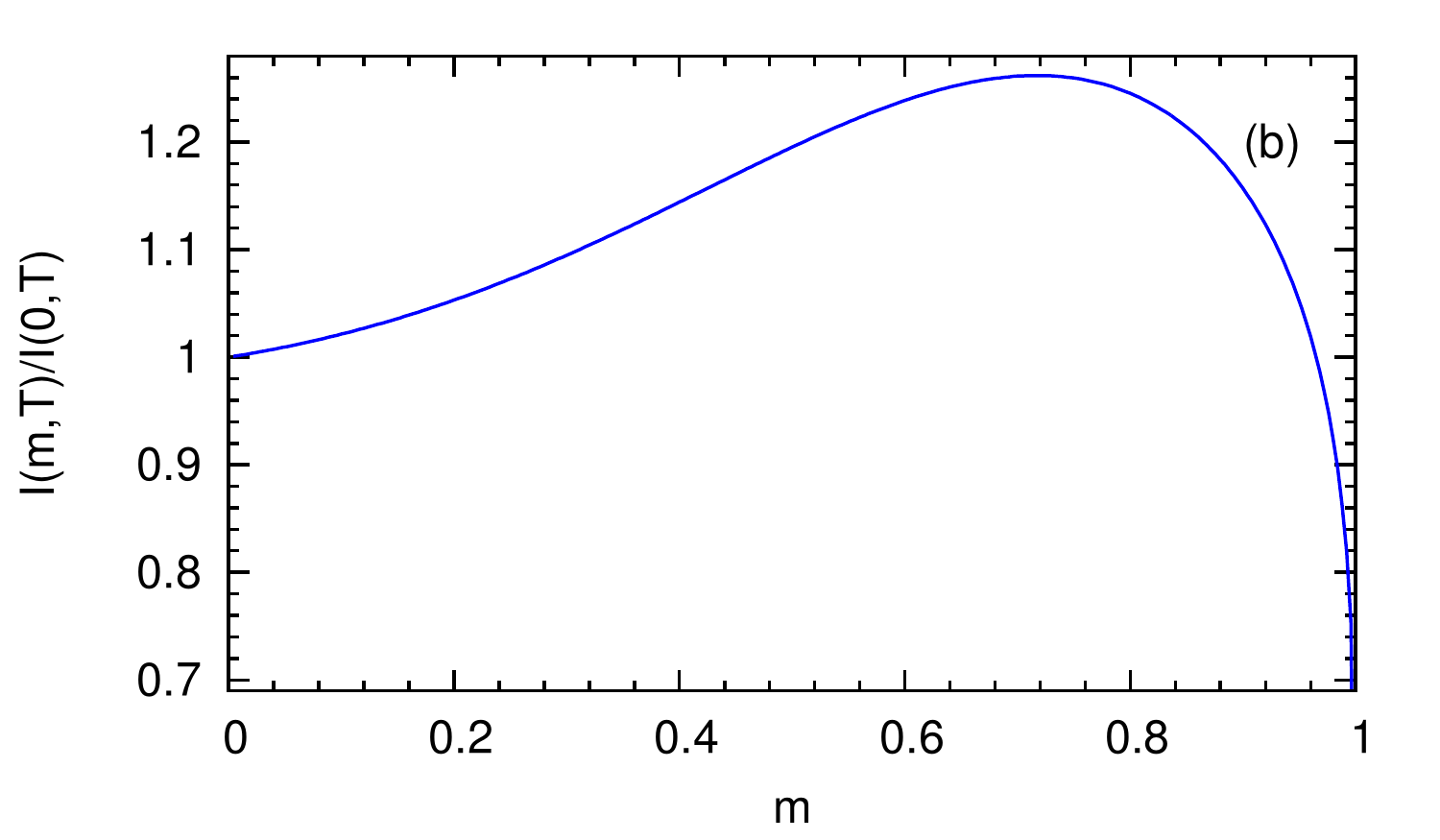} %
\includegraphics[width=0.6\columnwidth,angle=270]{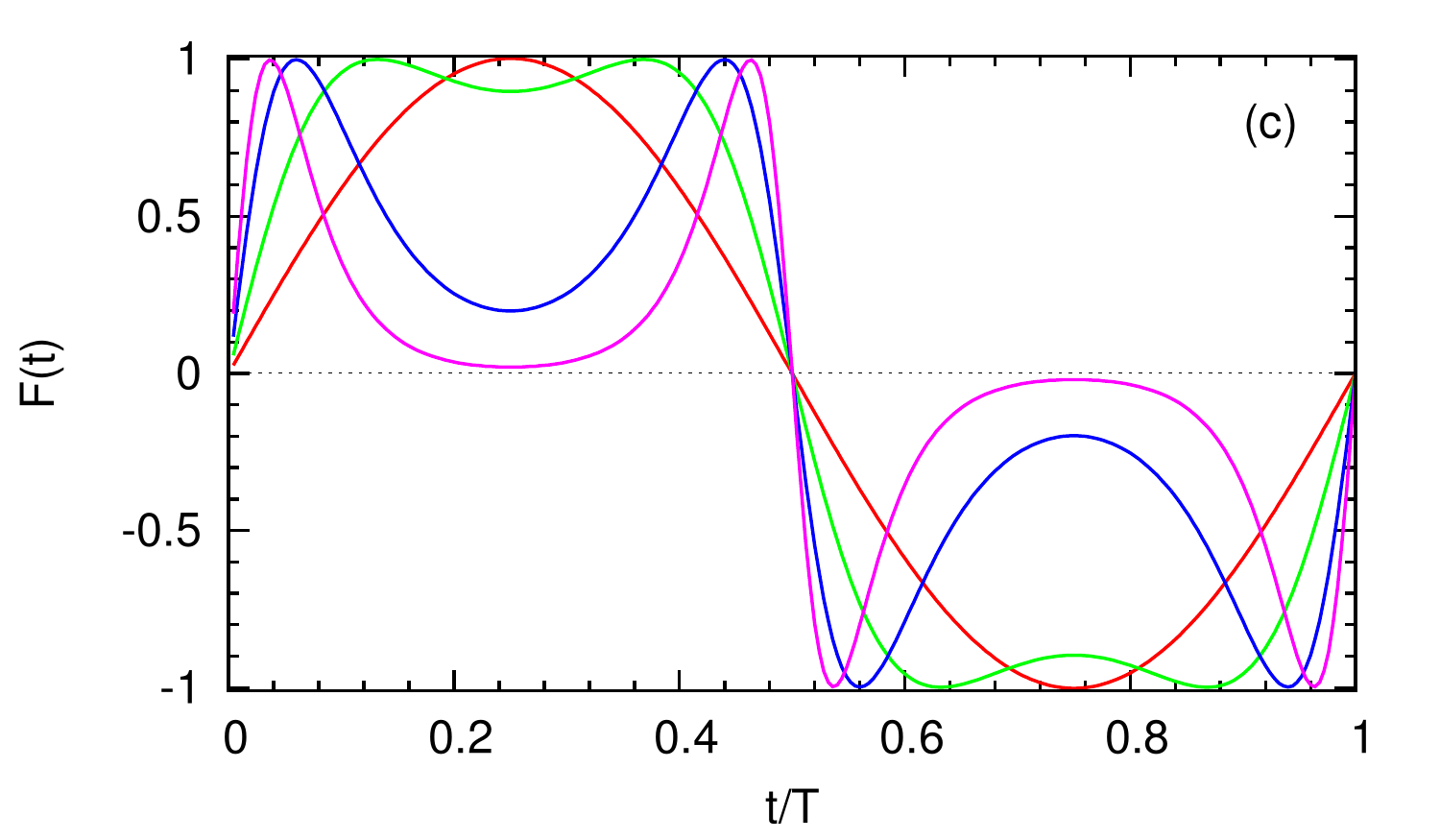} 
\caption{(a)~Normalization function $N(m)$, given by Eq.~\eqref{eq.N(m)},
  \textit{vs.}\ $m$. 
  (b)~Normalized force impulse, $I(m,T)/I(0,T)\equiv N(m)K(0)/[N(0)K(m)]$, 
  given by~Eq.~\eqref{eq.2}, \textit{vs.}\ $m$. 
  (c)~Force $F(t)$, given by Eq.~\eqref{eq.F}, \textit{vs.}\ $t/T$, 
  being~$T$ the period, for four values of the shape parameter: 
  $m=0$ (red sinusoidal pulse), $m=0.72$ (green nearly square-wave pulse), 
  $m=0.99$ (blue double-humped pulse), and 
  $m=0.999999$ (pink sharp double-humped pulse).}
\label{fig.1}
\end{figure}

The dynamics of our model system of ultracold atoms interacting with a
phase-modulated light field produced using an oscillating mirror is well
described by the periodic Hamiltonian 
%
\begin{equation}
\widetilde{H}=\widetilde{p}^{2}/(2M)- V_{0}\cos\left[ 2k\widetilde{x}-
\lambda F\left( t\right) \right],  
 \label{eq.Htilde}
\end{equation}
where $M$ is the atomic mass,~$\widetilde{x}$ the position,~$\widetilde{p}$
the momentum,~$V_0$ the potential height,~$k$ the wave number,~$\lambda$
the dimensionless modulation depth, and $F\left( t\right)$ the AC force
given by 
%
\begin{equation}
  F(t) =F( t;m,T) \equiv N(m) \; 
  \limfunc{sn}\left[ \frac{4Kt}{T}\right] 
  \limfunc{dn}\left[ \frac{4Kt}{T}\right],
\label{eq.F}
\end{equation}
where $\limfunc{sn}\left( \cdot\right) \equiv \limfunc{sn}\left(
\cdot;m\right)$ and $\limfunc{dn}\left( \cdot\right) \equiv \limfunc{dn}\left(
\cdot;m\right)$ are Jacobian elliptic functions of parameter $m$, and 
$K\equiv K(m)$ is the complete elliptic integral of the first kind. 
$N(m)$ is the normalization factor shown in Fig.~\ref{fig.1}(a), and given by 
%
\begin{equation}
  N(m) \equiv \frac{1}{a+\displaystyle\frac{b}{1+ \exp[(m-c)/d\;]}},
   \label{eq.N(m)}
\end{equation}
where the values of the parameters are set equal to $a\equiv0.43932$, 
$b\equiv0.69796$, $c\equiv0.3727$, and~$d\equiv0.26883$, 
in order to have the same force amplitude (equal to unity) and period, $T$, 
independently of the waveform, i.e. $\forall m\in\left[0,1\right]$, 
as shown in Fig.~\ref{fig.1}(c). 
For $m=0$, one recovers the well known harmonic excitation case
previously considered for example in Refs.~\onlinecite{5,6,7}, since 
then~$F\left( t;m=0,T\right) =\sin\left( 2\pi t/T\right)$. 
On the other hand, for $m \ne 0$ the waveform has different shapes. 
For example, for $m=0.72$ a nearly square--wave pulse is obtained, 
whereas for the limiting value $m=1$ the force vanishes.

As will be shown below, the force impulse associated with~$F(t)$, 
defined as 
%
\begin{equation}
  I\equiv I(m,T) =\int_0^{T/2} F(t;m,T) dt = \frac{TN(m)}{2K(m)} ,  
  \label{eq.2}
\end{equation}
is a relevant quantity to characterize the effect of the force's waveform.
As expected, it is a function of $m$, which has a single maximum at 
$m=m_{\max}^I\simeq 0.717$, as shown in Fig.~\ref{fig.1}(b). 
Also, it tends to zero very quickly as $m\rightarrow 1$. 
%

It is important to remark that our model
accurately describes a system of ultracold atoms in a periodically shaken
optical lattice in the absence of a ratchet effect, i.e.~with no directed
transport due to symmetry breaking of zero-mean forces~\cite{11}.
Nevertheless, it has been recently shown that optimum enhancement of ratchet
transport is achieved when maximal effective (critical) symmetry breaking
occurs, which is in turn a consequence of two reshaping-induced competing
effects: (i) the increase in the degree of symmetry breaking, and (ii) the
decrease in the (normalized) transmitted force impulse~\cite{12,13},
thus confirming the general relevance of this quantity.

Switching now to scaled dimensionless variables 
%
\begin{equation}
  \tau \equiv \omega t,\quad x\equiv 2k\widetilde{x},\quad 
  p\equiv \frac{2k}{M \omega}\; \widetilde{p},
\end{equation}
the Hamiltonian in Eq.~\eqref{eq.Htilde} is transformed into the following
dimensionless one 
%
\begin{equation}
  H\equiv \frac{4k^2}{M\omega^2}\widetilde{H} = \frac{p^2}{2}
  -\kappa \cos [x-\lambda F( \tau ;m) ] ,  
  \label{eq.3}
\end{equation}
where $\kappa \equiv V_{0}k^{2}T^{2}/\left( \pi ^{2}M\right) $, 
$F(\tau;m) \equiv N(m)\;\limfunc{sn}(\Omega\tau)\;\limfunc{dn}(\Omega \tau)$, 
and $\Omega \equiv 2K(m)/\pi $. 
By rewriting Eq.~\eqref{eq.3} in the form $H=H_{0}+H_{1}$, with 
%
\begin{align}
  H_{0}& \equiv \frac{p^2}{2}-\kappa \cos x, \\
  H_{1}& \equiv \kappa \left\{ \cos x-\cos \left[ x-\lambda N\left( m\right) 
  \limfunc{sn}\left( \Omega \tau \right) \limfunc{dn}\left( \Omega \tau
  \right) \right] \right\} ,
\end{align}
$H$ can be regarded now as the Hamiltonian
of an effective perturbed pendulum, being $H_{1}$ the perturbation
for~$\lambda~>~0$, such that expression 
%
\begin{equation}
  \frac{dH_{0}}{d\tau }=\kappa p\left\{ \sin x-\sin \left[ x
  -\lambda N\left(m\right) \limfunc{sn}\left( \Omega \tau \right) 
  \limfunc{dn}\left( \Omega \tau \right) \right] \right\}
\end{equation}
accounts for the effect of the perturbation on $H_{0}$. 
It has been shown for the cases of sinusoidal \cite{5,6} and 
square-wave \cite{10} forces that the strength of the DL 
of system \eqref{eq.Htilde} is correlated with the chaotic 
layer width of the aforementioned perturbed pendulum. 
For the present case, one can obtain an analytical estimate of the 
chaotic layer width for the case of small modulation amplitude such that 
%
\begin{equation}
\sin \left[ x-\lambda F\left( \tau ;m\right) \right] \approx 
  \sin x-\lambda F\left( \tau ;m\right) \cos x
\end{equation}
by calculating the change of the averaged energy along the separatrix 
%
\begin{align}
  x_{0,\pm }\left( \tau \right) & =\pm 2\arctan \left\{\sinh 
 \left[ \sqrt{\kappa }\left( \tau -\tau _{0}\right) \right] \right\},  \notag \\
  p_{0,\pm }\left( \tau \right) & =\pm 2\sqrt{\kappa }\func{sech}\left[ 
  \sqrt{\kappa }\left( \tau -\tau _{0}\right) \right]
\end{align}
of the unperturbed pendulum $H_{0}$ 
%
\begin{equation}
  \left\langle \frac{dH_{0}}{d\tau }\right\rangle \equiv 
  \int_{-\infty}^{\infty }\frac{dH_{0}}{d\tau }\;d\tau ,
\end{equation}
where $\tau =\tau _{0}$ is an arbitrary initial time~\cite{14}. 
After some simple algebraic manipulation, one straightforwardly obtains 
%
\begin{align}
  \max_{\tau _{0}}\left\langle dH_{0}/d\tau \right\rangle & \approx 
  d\left(\lambda ,\kappa ,m\right) 
  +\mathcal{O}\left( \lambda ^{2}\right), \notag \\
  d\equiv d\left( \lambda ,\kappa ,m\right) & =
  \frac{4\pi ^{3}\lambda N(m)}{\kappa \sqrt{m}K^{2}(m)}
  \sum_{n=0}^{\infty }a_{n}\left( \kappa \right)b_{n}\left( m\right) ,  
  \label{eq.d}
\end{align}
where 
%
\begin{align}
  a_{n}\left( \kappa \right) & \equiv \left( n+1/2\right)^{3}
  \func{sech}\frac{( n+1/2) \pi}{\sqrt{\kappa }} ,  \notag \\
  b_{n}\left( m\right) & \equiv \func{sech}\; 
  \frac{(n+1/2) \pi K(1-m)}{K(m)}.
\end{align}
The width function $d\left( \lambda ,\kappa ,m\right) $ gives a first-order
approximation in $\lambda $ to the width in energy of the chaotic separatrix
layer. 
As a function of $m$, it presents a single maximum at $m=m_{\max}^{d}\simeq 0.651$, 
which is significantly near $m_{\max }^{I}\simeq 0.717$,
in the sense that the respective wave forms are almost coincident 
(see discussion in Sec.~\ref{subsec.calc} below). 
Also, $\lim_{m\rightarrow 1}d\left(\lambda,\kappa,m\right)=0$, 
as the force vanishes in that limit \textit{irrespective} of the 
(finite) value of $\lambda$. 
Thus, one would expect the dependence of the width function on the
shape parameter to remain approximately (qualitatively) valid for values 
of the dimensionless modulation depth which are beyond the first-order
perturbative regime.

\section{Numerical results}

\label{sec.results}

In this section we compare the classical and quantum dynamics of 
our system using different tools.
First, by calculating the momentum distributions, we identify the
classical footprint of DL in the quantum dynamics. 
Second, to further demonstrate the influence of the classical phase 
space structures in the quantum behavior of the system, 
we compare the (classical) Poincar\'e surfaces of section (PSOS) 
and the corresponding Husimi based quantum surfaces of section 
(QSOS) \cite{18}, obtaining an excellent concordance between the 
results yielded by both tools.

\subsection{Momentum distributions}

\label{subsec.calc}

%
\begin{figure}
\includegraphics[width=0.9\columnwidth]{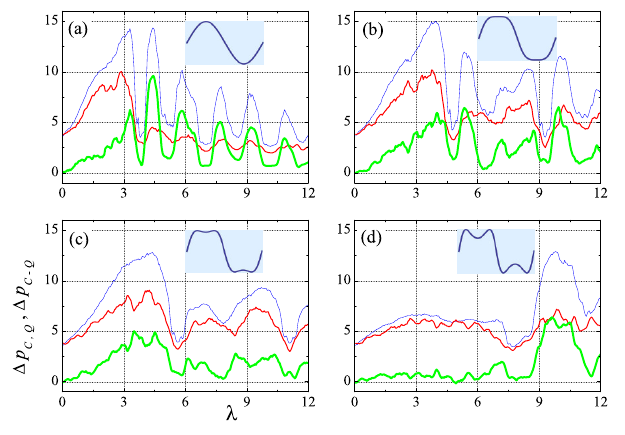}
\caption{Comparison of the classical (thin blue lines), $\Delta p_{C}$, and
quantum mechanical (medium red lines), $\Delta p_{Q}$, momentum
distributions of atoms, and their differences (thick green lines), 
$\Delta p_{C-Q}$, as functions of the modulation amplitude, $\protect\lambda $, 
for $\protect\kappa =0.36$, $\hbar_{\text{eff}}=0.16$ and four modulation
waveforms: (a) $m=0$, (b) $m=0.5$, (c) $m=0.7$, and (d) $m=0.9$. 
Insets: wave forms over a modulation period.}
\label{fig.2}
\end{figure}

The classical momentum distribution for our system is numerically 
calculated as 
%
\begin{equation}
P_{C}\left( p,\tau\right) =\int P\left( x,p;\tau\right) dx,
\end{equation}
where $P\left( x,p;\tau\right)$ is the evolution 
of a uniform distribution over one 
wavelength with a Gaussian momentum distribution characterized by 
a width of $\Delta p_{0}=0.386$~\cite{4,5,6} given by the 
corresponding Liouville equation, defined as
%
\begin{equation}
\left( \partial_{\tau}+p\partial_{x}-\kappa\sin\left[ x- \lambda
F\left(\tau;m\right)\right]\right)P\left(x,p;\tau\right)=0,
\end{equation}
subjected to periodic boundary conditions $P\left(x+L,p;\tau\right)
=P\left(x,p;\tau\right)$, $L=2n\pi$, with $L$ being the size of the
quantization box and $n\in \mathbb{Z}^{+}$.

On the other hand, quantum mechanical momentum distributions, 
$P_{Q}\left(p,\tau\right)$, were computed by averaging time 
propagations with the time-dependent Schr\"odinger equation 
%
\begin{equation}
i\hbar_\text{eff}\partial_{\tau}\psi = - \left\{\frac{\hbar_\text{eff}^2}{2}%
\partial_{xx}^{2} + \kappa\cos\left[ x-\lambda F\left(\tau;m\right)\right]%
\right\}\psi  \label{eq.13}
\end{equation}
of localized wave packets 
%
\begin{equation}
\psi\left( x,t\right) =\left( \pi\Delta x_{0}\right)^{-1/4} 
  \exp\left[-\frac{\left(x-x_0\right)^{2}}{2\Delta x_{0}}
  + \frac{ixp_{0}}{\hbar_\text{eff}}\right],  
  \label{eq.gauss}
\end{equation}
centered at $x_0\in\left[0,L\right]$ with $p_{0}=0$ and 
$\Delta x_{0}=\hbar_\text{eff}/\Delta p_{0}$, 
where $\hbar_\text{eff} \equiv 2\hbar k^2T/\left(\pi M\right)$ 
is the effective Planck constant characterizing the ``quanticity'' 
degree of the system.

The strength of DL is quantified by the difference between the 
classical and quantum mechanical momentum distributions 
$\Delta p_{C-Q}\equiv \Delta p_{C}-\Delta p_{Q}$, being 
%
\begin{equation}
\Delta p_{i}\equiv \frac{
  \sqrt{\left\langle p_i^2\right\rangle
  -\left\langle p_i\right\rangle^2}
  }{2\hbar_{\text{eff}}\ k},
 \qquad i=\left\{C,Q\right\},
\end{equation}
the normalized root-mean-square momentum widths. 
Recall that the larger $\Delta p_{C-Q}$, the stronger the DL.

In order to demonstrate that the force impulse is a relevant quantity
properly controlling DL, we show in Fig.~\ref{fig.2} the classical, 
$\Delta p_C$, and quantum, $\Delta p_Q$, momentum distributions as a 
function of $\lambda$, as well as the difference, $\Delta p_{C-Q}$, 
for four different values of $m$, i.e.~four different modulation waveforms. 
As it can be seen, the numerical results confirm that one needs larger 
values of the modulation amplitude $\lambda$ to obtain a noticeable DL 
as the force impulse (as a function of $m$) decreases from its maximum 
value at $m = m_\text{max}^I \simeq 0.717$.

%
\begin{figure}
\includegraphics[width=0.9\columnwidth]{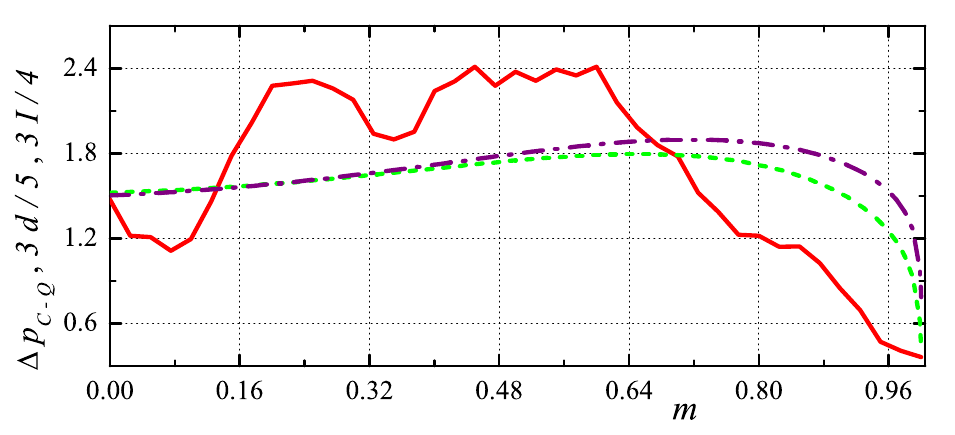}
\caption{Difference between the classical and quantum mechanical momentum
  distributions $\Delta p_{C-Q}$ (solid red line), 
  normalized width of the chaotic layer $3d/5$ given by Eq.~\eqref{eq.d} 
  (dashed green line), and the
  normalized force impulse $3I/4$ given by Eq.~\eqref{eq.2} 
  (dashed-dotted purple line) \textit{vs.} the shape parameter $m$. 
  The values of the fixed parameters are: $\protect\kappa =0.36$, 
  $\protect\lambda =2$, $\hbar_\textnormal{eff}=0.16$.}
\label{fig.3}
\end{figure}
In Fig.~\ref{fig.3}, we show the difference between the classical and the
quantum momentum distributions, $\Delta p_{C-Q}$, as a function of the 
shape parameter, $m$, keeping the rest of the parameters 
($\kappa $, $\lambda $, $\hbar_\text{eff}$) fixed. 
In this figure, we have superimposed the chaotic layer width in energy
\eqref{eq.2} and the force impulse. 
As it is shown, $\Delta p_{C-Q}$ presents its maximum values over a range 
$0.45\lesssim m\lesssim 0.6 $, which is close to the impulse maximum 
$m_{\max}^I$ [cf.~Fig.~\ref{fig.1}(b)]. 
Although not shown in Fig.~\ref{fig.3}, the three plotted quantities vanish in 
the limit $m\rightarrow 1$. 
Moreover, it is also important to remark the excellent agreement that 
exists between the chaotic layer width and the force impulse, 
that nearly coincide when adequately scaled. 
Thus, we can unambiguously conclude the correlation existing between these 
two quantities characterizing, respectively, the degree of chaoticity 
of the system and the AC force.

To further demonstrate the influence of the modulation waveform on DL, 
we show PSOSs for three different representative values of~$m$
in Fig.~\ref{fig.4} (top panel). 
These PSOSs have been stroboscopically computed at times multiples 
of the force period $T=2\pi $. 
One sees that the modulation waveform, and hence the force impulse,
does not only have a great influence 
on the the DL strength, but also on the phase space 
structures of the system.
Thus, for $m=0$ --a value of $m$ for which DL clearly occurs--  
there are only very tiny islands of regularity (leftmost panel), 
which are drastically reduced over the range of maximum strength of DL, 
as for $m=0.55$ for example (central panel). 
Finally, for values of $m$ which are sufficiently close to 1,
as for $m=0.9999$v(rightmost panel), DL is very weak, 
while the area occupied by regularity islands becomes even larger
as $m\rightarrow 1$. 
Clearly, this confirms the predicted DL scenario.

\subsection{Influence of the modulation waveform on the quantum dynamics}

\label{subsec.qsos}

%
\begin{figure*}[tbp]
\includegraphics[width=2\columnwidth]{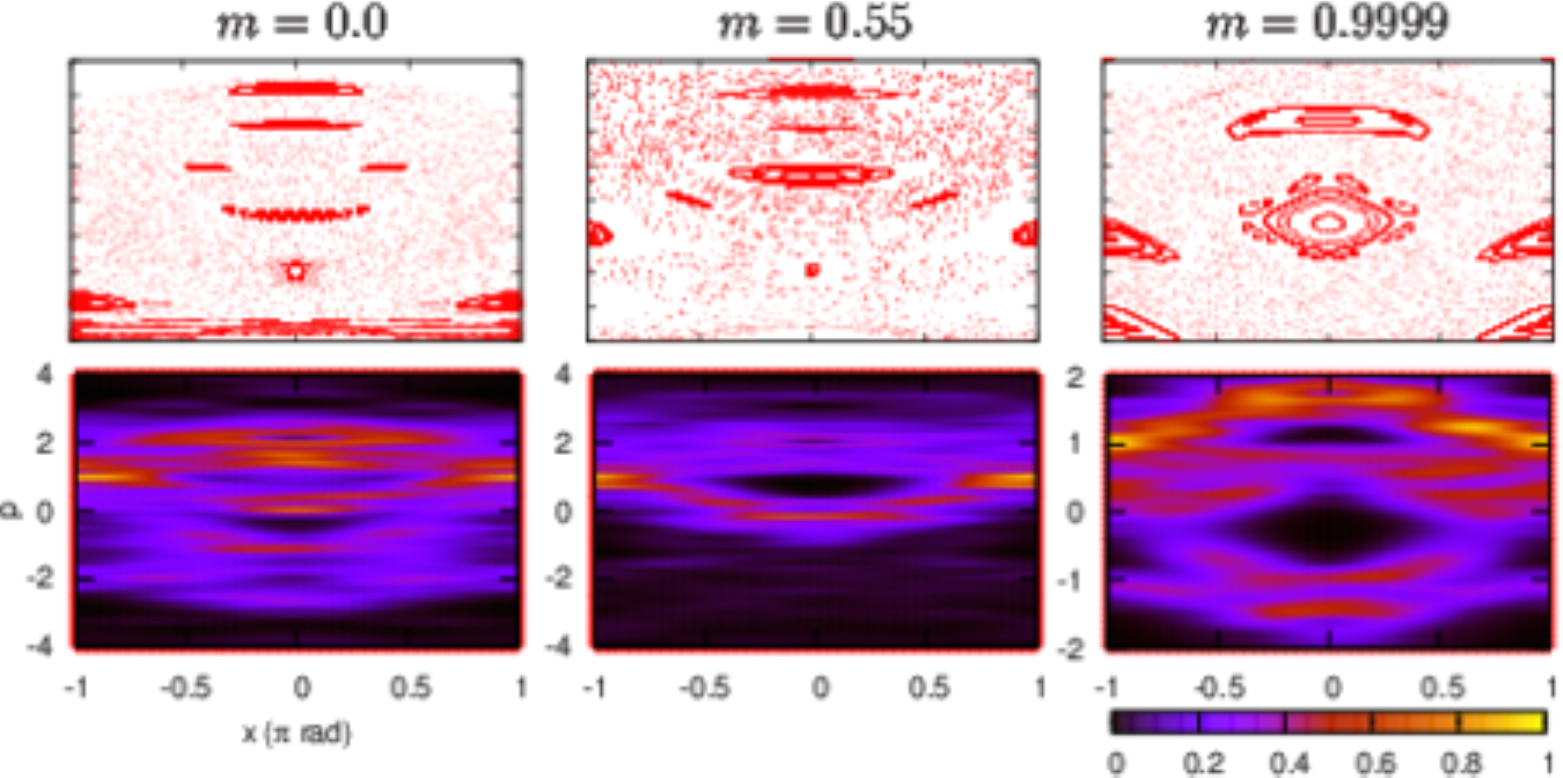}
\caption{Poincar\'e surfaces of section (top) and Husimi based quantum
  surfaces of section (bottom) obtained from the wave function given by 
  Eq.~\eqref{eq.gauss} with $x_0=\protect\pi$ and $p_0=1$ 
  averaged stroboscopically over 50 time periods for different values of $m$: 
  $m=0$ (left), $m=0.55$ (center), and $m=0.9999$ (right).}
\label{fig.4}
\end{figure*}
In this section we show how the quantum dynamics of our system, 
as it happens in general,
is dramatically influenced by the underlying classical structures and, 
as a consequence, by the modulation waveform. 
To demonstrate this effect, we calculate QSOSs for some representative 
values of the shape parameter $m$. 
Similarly to the classical PSOSs, we stroboscopically computed these QSOSs 
at multiples of the force period, $T=2\pi$, 
using the Husimi quasi-probability distribution \cite{18} 
\begin{widetext}
%
\begin{equation}
   {\cal H}(x,p,t)=\left\vert \; 
     \int_{-\infty}^\infty \exp\left\{-\frac{(x-z)^2}{2 \alpha}-
     i \frac{z p}{\hbar_\textrm{eff}} \right\} \psi(z,t)\; dz \;
     \right\vert^2 ,
  \label{eq.hus}
\end{equation}
with $\alpha=3$. 
This expression can be rewritten as a sum of infinite terms,
due to the existing periodic boundary conditions, as
%
\begin{equation}
   {\cal H}(x,p,t)=
  \left\vert \sum_{n=-\infty}^\infty \int_0^{2\pi} 
   \exp\left\{-\frac{[x-(z+2\pi n)]^2}{2 \alpha}-
   i \frac{(z+2\pi n) p}{\hbar_\textrm{eff}}\right\}
   \psi(z+2\pi n,t) \; dz 
   \; \right\vert^2, 
 \label{eq.hus2}
\end{equation}
although the calculation is effectively limited to the first three or four 
terms due to the gaussian decay.
\end{widetext} 
By averaging the QSOS over the times used in 
the calculation of the corresponding classical PSOS, 
we can quantum mechanically \textit{unfold} the phase 
space structure of the system, which consists of a mix of islands of 
regularity and regions of stochasticity. 
Moreover, this averaged QSOS can be directly compared with the 
corresponding classical PSOS.
Some results for different representative values of the parameter 
$m$ are shown in Fig.~\ref{fig.4} (bottom panels), 
where the QSOSs after averaging over 50 modulation periods for a coherent
state \eqref{eq.gauss} centered at ($x_0,p_0)=\pi,1$ are depicted.
This initial condition gives rise to a classically chaotic trajectory. 
One sees that, the three QSOSs appear all spread out over the 
corresponding chaotic regions of phase space. 
Also notice how the probability density does not enter into the 
regions where the motion is regular, since the corresponding tori 
and resonant islands act as total barriers both for the classical 
motion and the flow of quantum probability across.
This effect is clearly more noticeable as $m$ is increased, 
i.e.~when larger portions of phase space become regular. 
As a consequence, the QSOSs mimic to a great 
extend the corresponding classical PSOSs. 
One point is worth mentioning here.
If we had propagated the initial wave functions for a much longer time, 
the corresponding QSOSs would have entered to some extend inside 
these regions of regularity due to the usual quantum tunneling.

Let us consider now the quantum dynamics associated with other types
of classical motion.
When the wave function is initially localized at a point associated 
with a regular classical trajectory, whose dynamics is then confined 
to the corresponding invariant tori or chain of islands, 
the quantum quasiprobability density behaves accordingly, 
thus remaining well localized in the neighborhood of the tori 
(except, again, for very long times when quantum tunneling becomes 
noticeable). 
%
\begin{figure}[tbp]
\includegraphics[width=0.9\columnwidth]{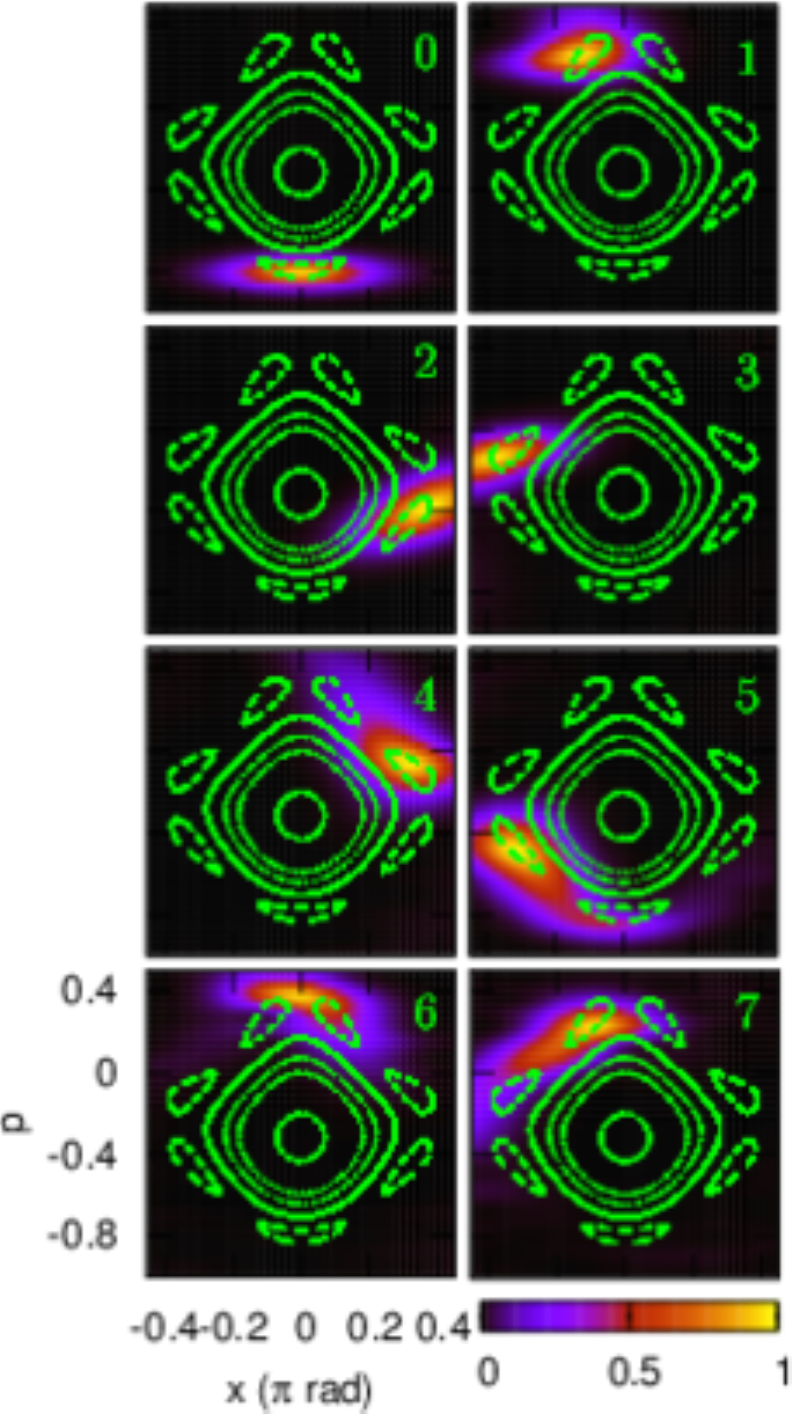}
\caption{Husimi based quantum surface of section \eqref{eq.hus2} 
  for the wave function~\eqref{eq.gauss} launched at $(x_0,p_0)=(0,-0.805)$ 
  for $m=0.9999$. 
  The corresponding classical Poincar\'e surface of section has
  been represented superimposed in green dots. The number at the 
  top right corner of each panel indicates the elapsed time in units 
  of the modulation period $T=2\pi$.}
\label{fig.5}
\end{figure}
An example is depicted in Fig.~\ref{fig.5}, where we show the
QSOS corresponding to a wave function for $m=0.9999$ and 
 $(x_0, p_0)=(0,-0.805)$. 
This initial condition generates a classical trajectory 
in the chain of islands associated with a 7:3 resonance 
(see the corresponding PSOS in Fig.~\ref{fig.5}). 
One sees that the localized wave function jumps over three
of the islands at the instants $t=T,2T,\ldots,7T$ and returns
close (although with some accumulated delay) to the 
initial position at $t=8T$.
Note that this is the behavior expected from the dynamics 
associated with the 7:3 classical resonance.

%
\begin{figure}[tbp]
\includegraphics[width=0.9\columnwidth]{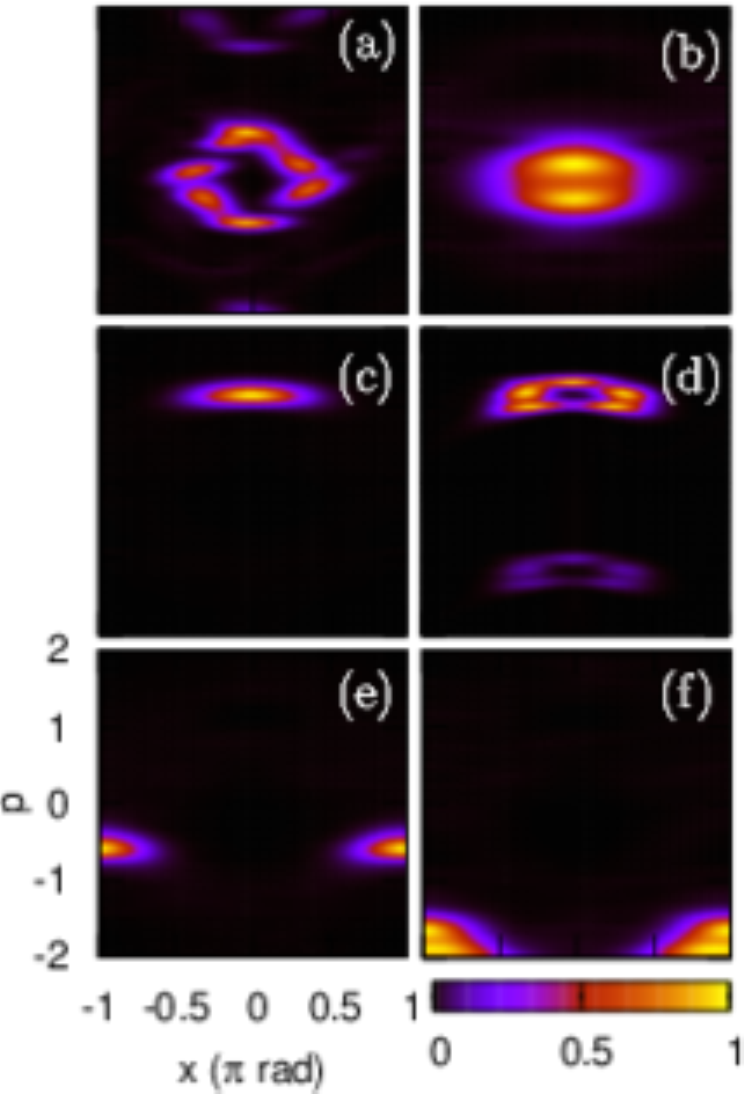}
\caption{Long-time averaged Husimi based quantum surfaces of section
corresponding to wave function~\eqref{eq.gauss} initially launched at
different points in phase space for $m=0.9999$.
All the panels have been (stroboscopically) averaged 
over~50$T$, except panels (a) and (d), 
which have been averaged over $7T$ and $5T$, respectively.}
\label{fig.6}
\end{figure}
It is worth noting that if we add all the QSOSs distributions shown 
in Fig.~\ref{fig.5} and renormalize the total, we get a new 
distribution highly localized on the corresponding classical PSOS. 
This total distribution is depicted in Fig.~\ref{fig.6}(a),
where the described effect is clearly visible.
Figures~\ref{fig.6}(b)-(f) show similar results for QSOS 
localized on other relevant regular structures existing in the classical 
PSOS of Fig.~\ref{fig.4} for $m=0.9999$.
For this purpose, we have used wave functions launched at 
suitable initial conditions in phase space, and we have 
(stroboscopically) averaged the obtained QSOS over~50$T$. 
One sees that these averaged QSOS appear respectively 
localized on a tori around the central fixed point $(x,p)=(0,0)$
[Fig.~6(b)], 
on a tori around the phase space point (0,2) [Fig.~6(c)],
on the 5:1 chain of islands around the same point [Fig.~6(d)],
on a tori around the border fixed point at ($\pm 1$,--0.5) [Fig.~6(e)],
and on a tori around the other border fixed point at 
($\pm 1$,--2) [Fig.~6(f)].
In all cases, the QSOSs completely mimic the structure 
of the classical phase space. 
This clearly demonstrates the importance of this classical structures 
for the quantum motion. 
It is worth mentioning that similar results to those presented here 
have been obtained for other values of the shape parameter $m$.

\section{Conclusions and final remarks}

\label{sec.conclu}

Summarizing, in this paper it has been shown for AC-driven space-periodic 
Hamiltonians that the impulse transmitted by the time-periodic force 
is an essential quantity to optimally control the wave-packet spreading 
associated with the phenomenon of dynamical localization. 
While this result holds for the wide class of AC forces having equidistant 
consecutive zeros, one expects it to be also valid for even more general 
class of time-periodic forces, at least in the adiabatic regime~\cite{15}. 
This principle, which can be straightforwardly applied to other phenomena, 
such as field-induced barrier transparency~\cite{16} or quasi-energy 
band collapse~\cite{17}, paves the way for optimum coherent control of diverse 
quantum systems. 
Additionally, it has been shown that the classical invariant structures 
of phase space have a deep impact on the quantum dynamical behavior 
of the system, as the force impulse is varied.


\section*{Acknowledgements}

This research has been supported by the Ministry of Economy and
Competitiveness (MECC, Spain) under Contract Nos.~MTM2012-39101, 
FIS2012-34902, and ICMAT Severo Ochoa~\mbox{SEV-2011-0087},
and Junta de Extremadura (JEx, Spain) though 
project N$^\circ$~GR10045.


\begin{thebibliography}{99}
\bibitem{1} A. Einstein, 
  Deutsche Physikalische Gesellschaft Verhandlungen \textbf{19}, 82 (1917).

\bibitem{2} G. Casati, B. V. Chirikov, F. M. Izrailev, and J. Ford, 
  Lect. Notes Phys. \textbf{93}, 334 (1979).

\bibitem{3} S. Fishman, D. R. Grempel, and R. E. Prange, 
  Phys. Rev. Lett. \textbf{49}, 509 (1982).

\bibitem{4} F. L. Moore, J. C. Robinson, C. Bharucha, P. E. Williams, 
  and M. G. Raizen, Phys. Rev. Lett. \textbf{73}, 2974 (1994).

\bibitem{5} R. Graham, M. Schlautmann, and P. Zoller, 
  Phys. Rev. A \textbf{45}, R19 (1992).

\bibitem{6} R. Chac\'{o}n, Europhys. Lett. \textbf{77}, 30001 (2007).

\bibitem{7} G. Abal, R. Donangelo, A. Romanelli, A. C. Sicardi-Schifino, 
  and R. Siri, Phys. Rev. E \textbf{65}, 046236 (2002).

\bibitem{8} L. E. Reichl, 
  \textit{The Transition to Chaos in Conservative Classical Systems: 
  Quantum Manifestations} (Springer-Verlag, New York, 1992).

\bibitem{9} G. Casati and B. V. Chirikov (eds), 
  \textit{Quantum Chaos between Order and Disorder} 
  (Cambridge University Press, Cambridge, 1995).

\bibitem{10} R. Chac\'{o}n, F. Borondo, and D. Farrelly, 
  Europhys. Lett. \textbf{86}, 30004 (2009).

\bibitem{11} P. H\"{a}nggi and F. Marchesoni, 
  Rev. Mod. Phys. \textbf{81}, 387 (2009).

\bibitem{12} R. Chac\'{o}n, J. Phys. A \textbf{40}, F413 (2007).

\bibitem{13} R. Chac\'{o}n, J. Phys. A \textbf{43}, 322001 (2010).

\bibitem{14} A. J. Lichtenberg and M. A. Lieberman, 
  \textit{Regular and Chaotic Dynamics} (Springer-Verlag, Berlin, 1991).

\bibitem{15} R. Chac\'{o}n, M. Yu. Uleysky, and D. V. Makarov, 
  Europhys. Lett. \textbf{90}, 40003 (2010).

\bibitem{16} I. Vorobeichik, R. Lefebvre, and N. Moiseyev, 
  Europhys. Lett. \textbf{41}, 111 (1998).

\bibitem{17} D. H. Dunlap and V. M. Kenkre, 
  Phys. Rev. B \textbf{34}, 3625 (1986).

\bibitem{18} K. Husimi, 
  Proc. Phys. Math. Soc. Jpn. \textbf{22}, 264 (1940).

\end{thebibliography}
\end{document}